\documentclass[pdflatex,sn-mathphys-num]{sn-jnl}

\usepackage{graphicx}%
\usepackage{multirow}%
\usepackage{amsmath,amssymb,amsfonts}%
\usepackage{amsthm}%
\usepackage{mathrsfs}%
\usepackage[title]{appendix}%
\usepackage{xcolor}%
\usepackage{textcomp}%
\usepackage{manyfoot}
\usepackage{booktabs}%
\usepackage{algorithm}%
\usepackage{algorithmicx}%
\usepackage{algpseudocode}%
\usepackage{listings}%
\usepackage{indentfirst}
\theoremstyle{thmstyleone}%
\theoremstyle{thmstyletwo}%

\theoremstyle{thmstylethree}%

\begin{document}

\title[Article Title]{Search for the QCD Critical Point in High Energy Nuclear Collisions: A Status Report}

\author[1]{\fnm{Yu} \sur{Zhang}}\email{zhangyu@gxnu.edu.cn}
\author[2]{\fnm{Zhaohui} \sur{Wang}}\email{zh-wang@mails.ccnu.edu.cn}
\author*[2]{\fnm{Xiaofeng} \sur{Luo}}\email{xfluo@mail.ccnu.edu.cn}
\author*[2]{\fnm{Nu} \sur{Xu}}\email{nuxu02@gmail.com}

\affil[1]{\orgdiv{School of Physical Science and Technology}, \orgname{Guangxi Normal University}, \orgaddress{\street{15 Yucai Road}, \city{Guilin}, \postcode{541004}, \country{China}}}

\affil[2]{\orgdiv{Key Laboratory of Quark \& Lepton Physics (MOE) and Institute of Particle Physics}, \orgname{Central China Normal University}, \orgaddress{\street{152 Luoyu Road}, \city{Wuhan}, \postcode{430079}, \country{China}}}

\abstract{We review recent results of net-proton multiplicity fluctuations from STAR experiment, aiming to locate the QCD critical point in high-energy nuclear collisions at RHIC. We show net-proton number cumulant and proton number factorial cumulant ratios up to fourth order using experimental data from RHIC BES-II Au+Au collisions in collider mode and fixed-target mode. 
The comparison is made between experimental data and 
non-critical model calculations from Lattice QCD, HRG, hydrodynamic simulations and transport model UrQMD. In addition, we discuss initial volume fluctuation effect, which plays significant role in fixed-target energies. Finally, an outlook on experimental research on the QCD critical point in future experiments will be presented.}

\maketitle

\section{Introduction}\label{sec:intro}

One of main goals in heavy-ion collisions is to explore the phase structure of strongly interacting matter.
In the conjectured QCD phase diagram, plotted in terms of temperature $\rm T$ and baryon chemical potential $\mu_{\rm B}$, lattice QCD calculations from first principles predict a smooth crossover transition between the quark–gluon plasma (QGP) and hadronic phase at vanishing $\mu_{\rm B}$~\cite{Aoki:2006we}. However, at large $\mu_{\rm B}$, lattice QCD encounters the fermion sign problem and cannot provide definitive conclusions about the nature of the QCD phase transition.
\begin{figure}[htb]
\centering
\includegraphics[width=1\textwidth]{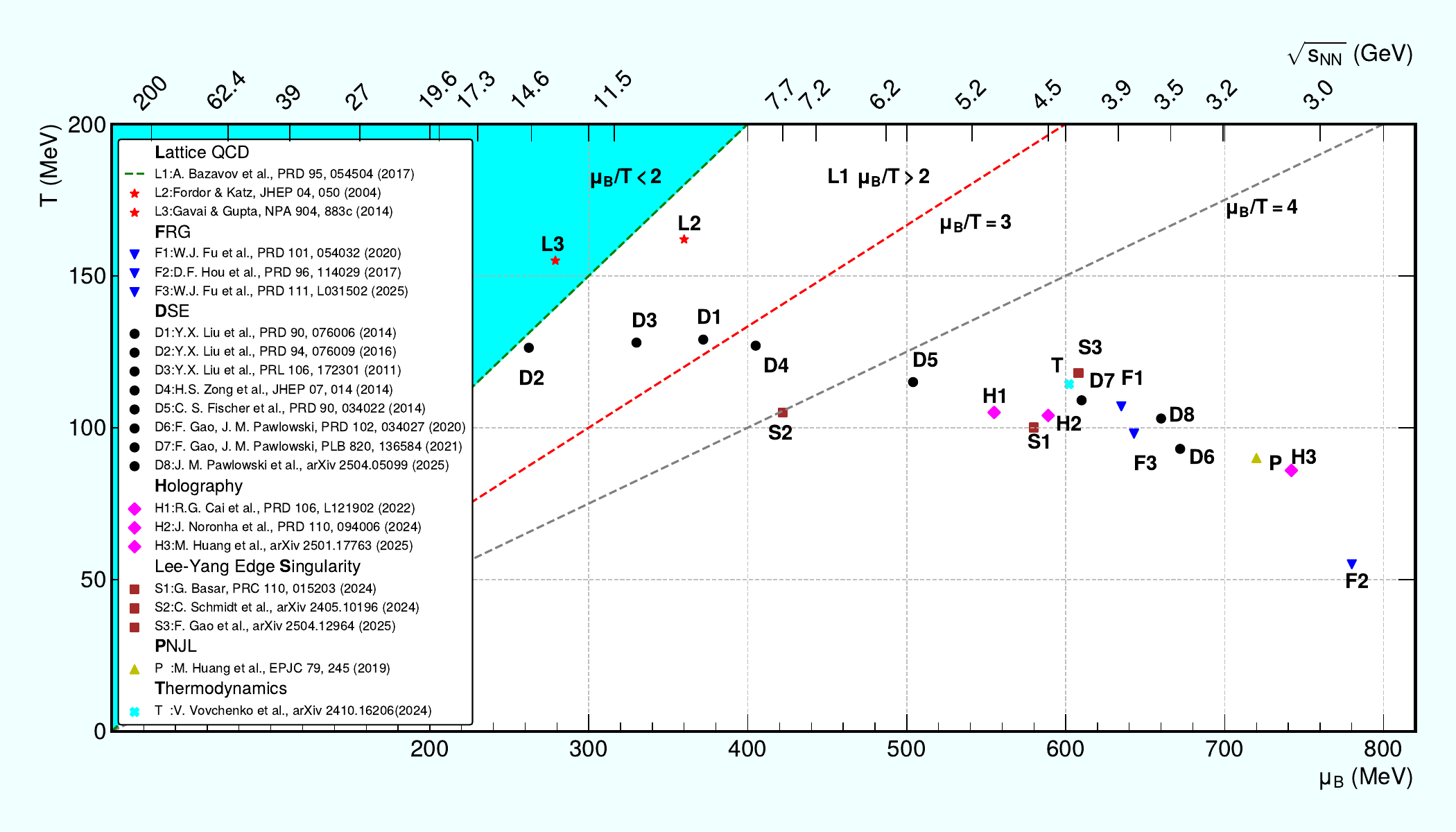}
\caption{Theoretical predictions on the QCD critical point from various approaches.}
\label{fig:cp}
\end{figure}

Many theoretical calculations predict that, at high baryon chemical potential, the transition between the hadronic phase and the QGP phase becomes first-order, terminating at a second-order phase transition point commonly referred to as the QCD critical point. On the experimental side, extensive measurements of various observables have been carried out in an effort to locate this critical point. Among these, higher-order cumulants of conserved quantities, such as net baryon number, net electric charge, and net strangeness, have been shown to be particularly sensitive probes of critical fluctuations~\cite{Stephanov:2008qz,Asakawa:2009aj,Stephanov:2011pb,Bzdak:2019pkr,Luo:2017faz,Pandav:2022xxx,Braun-Munzinger:2026krf}. A key signature of the QCD critical point is the non-monotonic energy dependence of the fourth-order cumulant ratio ($\kappa\sigma^{2}$)~\cite{Stephanov:2008qz}. Such measurements have been performed by the STAR Collaboration at the Relativistic Heavy Ion Collider (RHIC)~\cite{STAR:2002eio,STAR:2010mib,STAR:2013gus,STAR:2020tga,STAR:2021iop,STAR:2021rls,STAR:2022vlo,STAR:2021fge,STAR:2022etb}. 
While fluctuations of conserved quantities are sensitive to critical phenomena, their interpretation in heavy-ion collisions requires careful consideration of finite-size effects, critical slowing down, and non-equilibrium dynamics such as baryon transport. These effects are not merely additive to equilibrium baselines but can fundamentally alter fluctuation observables. Therefore, deviations from model calculations should be interpreted with caution, and alternative approaches incorporating finite-size scaling~\cite{Fraga:2011hi, Sorensen:2020ygf,Sorensen:2024mry,Fu:2019hdw,Gao:2020fbl,Gunkel:2021oya} and dynamical modeling are also essential in the search for the QCD critical point.
Recently, theoretical studies~\cite{Fu:2023lcm,Lu:2025cls,Cai:2022omk,Hippert:2023bel,Zhu:2025gxo,Basar:2023nkp,Clarke:2024ugt,Wan:2025wdg,Li:2018ygx,Shah:2024img,Jokela:2024xgz,Ecker:2025vnb} employing diverse approaches have converged on a critical point located in the high, $\mu_{\rm B}$ region (see Fig.~\ref{fig:cp}), a domain accessible to the RHIC Beam Energy Scan program~\cite{STAR:2017sal,Chen:2024aom} and the STAR fixed-target experiment~\cite{Meehan:2016iyt}.

This review is organized as follows. In Sec.~\ref{sec:exp}, we briefly introduce the RHIC-STAR experiment and details of higher-order cumulant measurements thereon. Recent progress on (net-)proton multiplicity fluctuation analyses from STAR is presented in Sec.~\ref{sec:result}, and we conclude with a summary in Sec.~\ref{sec:summary}.

\section{Experiment and Data Analysis}\label{sec:exp}
During the Beam Energy Scan (BES) program at RHIC from 2010 to 2021, the STAR detector collected a set of gold on gold nucleus collision data across a center-of-mass energy range of $\sqrt{s_{\rm NN}}=3-200$ GeV, operating at collider mode and fixed-target mode (FXT). The Au+Au collision data span a broad range of $\mu_{\rm B}$ $\approx$ 25-750 MeV at chemical freeze-out~\cite{STAR:2017sal}. The measurements emphasized in this review utilize data from the second phase of the BES program (BES-II), which includes collider-mode data at $\sqrt{s_{\rm NN}}=7.7$, 9.2, 11.5, 14.6, 17.3, 19.6, and 27 GeV, as well as fixed-target data at $\sqrt{s_{\rm NN}}=3.0$, 3.2, 3.5, and 3.9 GeV.

The observables used to characterize (net-)proton multiplicity fluctuations are $n$-th order cumulants ($C_n$) and factorial cumulants ($\kappa_n$) in Eq.~(\ref{eq:observable}), where $N$ denotes (net-)proton number. Proton number is used as a proxy for baryon number as electrically neutral particles, mainly neutrons in baryon fluctuation analysis, cannot be detected by STAR detector. We present the direct measurements of net-proton fluctuations in this review. As claimed in Refs.~\cite{Kitazawa:2011wh, Kitazawa:2012at}, over a wide energy range, isospins of nucleons are randomized and uncorrelated in the final state due to reactions of nucleons with pions in the hadronic stage, which leads to a factorization of proton and neutron probability distributions. In this case, net-baryon cumulants can be obtained in terms of net-proton cumulants.

The deviation of $N$ from its mean value $\langle N\rangle$ is defined as $\delta N=N-\langle N\rangle$. Ratios of cumulants ($C_n$) or factorial cumulants ($\kappa_n$) of different orders are constructed in order to cancel volume dependence and to enable direct comparison with susceptibility ratios from theoretical calculations~\cite{Gavai:2010zn,Gupta:2011wh,Karsch:2010ck,Garg:2013ata}.

\begin{align}\label{eq:observable}
\begin{aligned}
    C_{1} &= \langle N\rangle, & 
    \kappa_{1} &= C_{1}, \\
    C_{2} &= \langle (\delta N)^{2}\rangle, & 
    \kappa_{2} &= -C_{1}+C_{2}, \\
    C_{3} &= \langle (\delta N)^{3}\rangle, & 
    \kappa_{3} &= 2C_{1}-3C_{2}+C_{3}, \\
    C_{4} &= \langle (\delta N)^{4}\rangle-3\langle (\delta N)^{2}\rangle^{2}, & 
    \kappa_{4} &= -6C_{1}+11C_{2}-6C_{3}+C_{4}.
\end{aligned}
\end{align}

The event-by-event (net-)proton multiplicity distributions are constructed by identifying (anti-)protons using STAR time projection chamber (TPC)~\cite{Anderson:2003ur} and time-of-flight (TOF)~\cite{Llope:2012zz} detectors. Combining trajectory information from the TPC with flight-time measurements from the TOF enables proton identification with a purity exceeding 99\%. Collision centrality is determined indirectly using the multiplicity distribution of charged hadrons excluding (anti-)protons. These reference multiplicity distributions are then compared with Monte Carlo Glauber simulations~\cite{Miller:2007ri} and classified into centrality bins such as 0–5\%, 5–10\%, ..., 70–80\%.

Thanks to the upgrade of inner time projection chamber (iTPC)~\cite{star:itpc} to STAR TPC, the reference multiplicity distributions in BES-II($\sqrt{s_{\rm NN}}=7.7-$19.6 GeV) is measured over an extended pseudo-rapidity acceptance $|\eta|<1.6$, compared to $|\eta|<1.0$ in BES-I. Two other major upgrades, namely end-cap time of flight (eTOF)~\cite{STAR:2016gpu} and event plane detector (EPD)~\cite{Adams:2019fpo} have further enhanced STAR’s capabilities. Particularly, the eTOF detector upgrade plays a crucial role in particle identification for the STAR fixed-target experiment. 

Cumulants of (net-)proton multiplicity distributions are calculated in fine reference multiplicity intervals and then averaged over broader centrality classes (e.g. 0-5\%) following the centrality bin width correction procedure shown in Ref.~\cite{Luo:2013bmi}. Analytical corrections for detector efficiency loss~\cite{Nonaka:2017kko,Luo:2018ofd} are applied under the assumption of a binomial detector response~\cite{STAR:2013gus,Luo:2014rea}. Additional experimental factors, such as track selection criteria, are considered in systematical uncertainty evaluation.

\section{Recent results}\label{sec:result}
\subsection{(Net-)proton multiplicity fluctuations from STAR}
\begin{figure}[htb]
\includegraphics[width=0.99\textwidth]{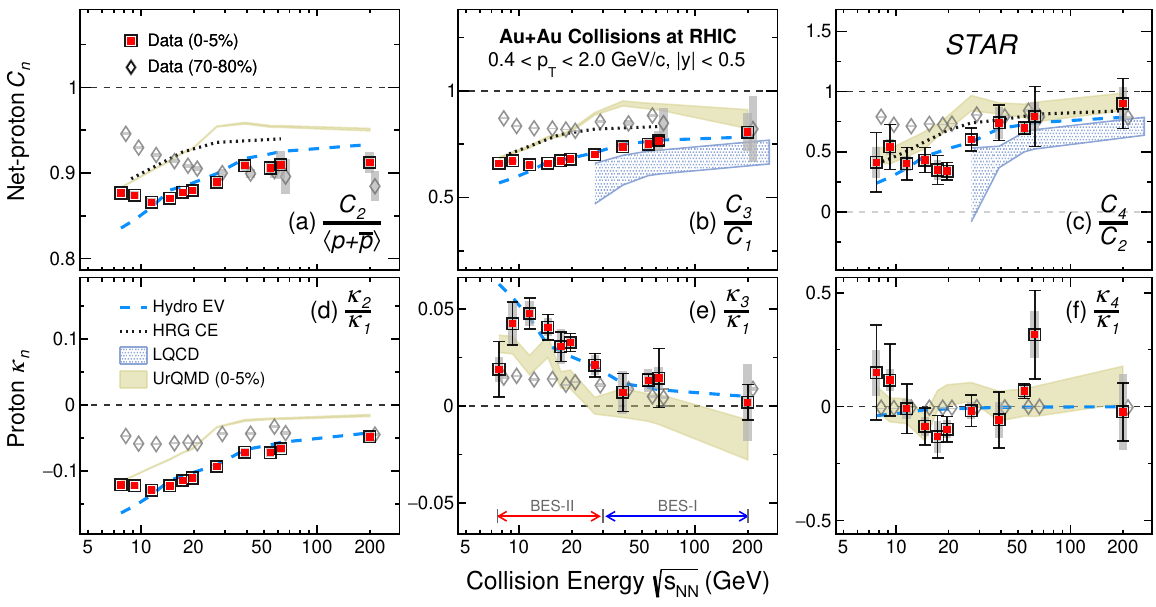}
\centering
\caption{Net-proton cumulant ratios: (a) $C_2/\langle p+\bar{p}\rangle$, (b) $C_3/C_1$, and (c) $C_4/C_2$ and proton factorial cumulant ratios: (d) $\kappa_2/\kappa_1$, (e) $\kappa_3/\kappa_1$, and (f) $\kappa_4/\kappa_1$ in Au+Au collisions from STAR at RHIC. Figure is taken from Reference~\cite{STAR:2025zdq}.}
\label{fig:BESII_C4}
\end{figure}

Figure~\ref{fig:BESII_C4} shows collision energy dependence of net-proton cumulant ratios $C_{2}/\langle p+\bar{p}\rangle$, $C_{3}/C_{1}$, and $C_{4}/C_{2}$ and proton factorial cumulant ratios $\kappa_2/\kappa_1$, $\kappa_3/\kappa_1$, and $\kappa_4/\kappa_1$ in Au+Au collisions from BES-II ($\sqrt{s_{\rm NN}}=7.7-27$ GeV)~\cite{STAR:2025zdq} and BES-I ($\sqrt{s_{\rm NN}}=39-200$ GeV)~\cite{STAR:2020tga,STAR:2021iop} at RHIC. Measurements are performed in 0-5\% most central (red filled squares) and 70-80\% peripheral collisions (black open diamonds) with the rapidity window $|y|<0.5$ in the center-of-mass frame and transverse momentum range $0.4<p_{\rm T}<2.0$ GeV/$c$. The bars and bands on the data points reflect statistical and systematical uncertainties, respectively. Theoretical calculations from various approaches are also presented for comparison: a hydrodynamical model with repulsive interactions via excluded-volume corrections~\cite{Vovchenko:2021kxx} (Hydro EV, blue dashed line), the thermal model with canonical treatment for baryon number~\cite{Braun-Munzinger:2020jbk} (HRG CE, black dotted line), 
the UrQMD transport model~\cite{Bass:1998ca,Bleicher:1999xi} (brown band), and lattice QCD calculations for net-baryon number fluctuations~\cite{Bazavov:2020bjn,Bollweg:2024epj} (light blue dotted band).

\begin{figure}[htb]
\centering
\includegraphics[width=0.55\textwidth]{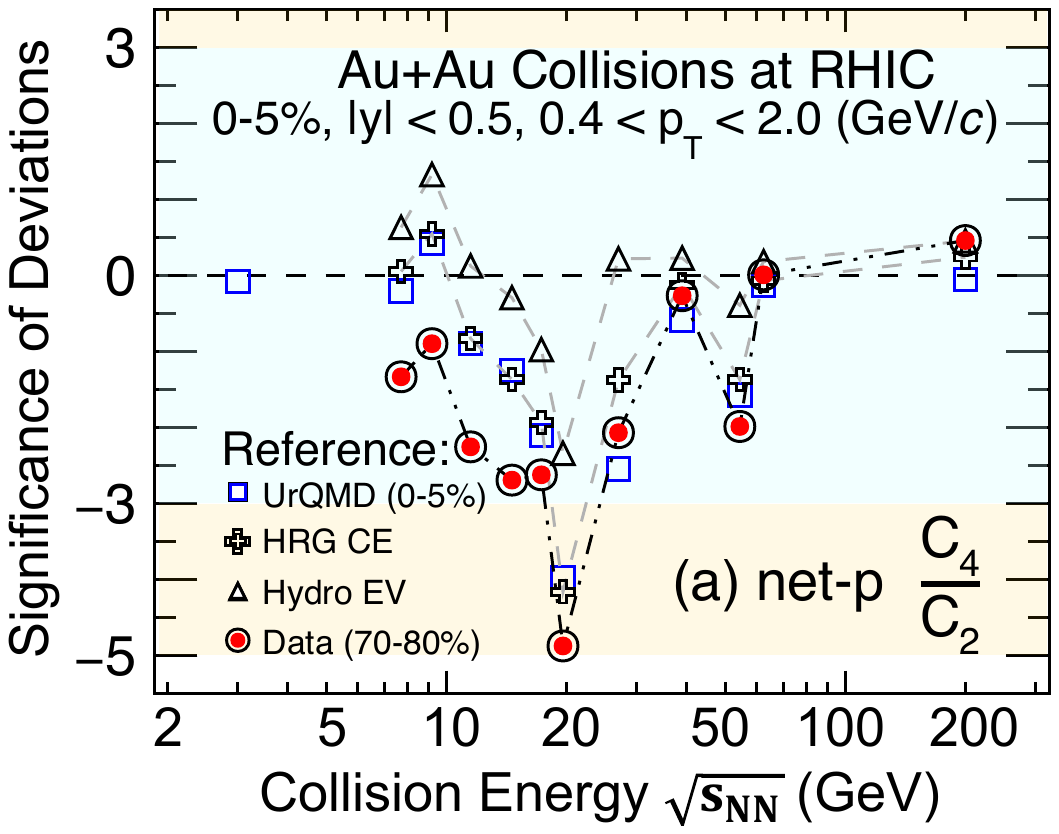}
\caption{Significance of deviation, (data$-$reference)/$\sigma_{\rm total}$, for (a) net-proton cumulant ratios $C_4/C_2$ in 0-5\% Au+Au collisions~\cite{STAR:2020tga,STAR:2021iop,STAR:2021fge,STAR:2022etb}. References include the non-critical model calculations, such as the UrQMD transport model~\cite{Bass:1998ca,Bleicher:1999xi} (blue square), HRG with canonical ensemble for baryon charge~\cite{Braun-Munzinger:2020jbk} (HRG CE, black cross), hydrodynamic model with excluded volume~\cite{Vovchenko:2021kxx} (Hydro EV, black triangle), and data from 70-80\% peripheral collisions (red dots). Figure is taken from Reference~\cite{STAR:2025zdq}.}
\label{fig:significance}
\end{figure}

As the energy decreases from 200 GeV, net-proton $C_2/\langle p+\bar{p}\rangle$, $C_3/C_1$, and $C_4/C_2$ in 0-5\% central collisions show monotonically decreasing trends which can also be seen in theoretical calculations. At high energies, $\sqrt{s_{\rm NN}} > 40$ GeV, the decreasing trends are reproduced by model calculations including Lattice QCD, HRG and hadronic transport model UrQMD. While when it goes below around 11 GeV, all ratios show clear increase trends especially for $C_2/\langle p+\bar{p}\rangle$ and $C_3/C_1$, which could not be reproduced by theoretical calculations. To quantify the deviations of net-proton $C_{4}/C_{2}$ of data from non-critical baselines, significance (shown in Fig.~\ref{fig:significance}) are calculated by (data - reference) divided by total uncertainties. The non-critical references include UrQMD, HRG CE, Hydro EV, and data from 70-80\% peripheral collisions. From Fig.~\ref{fig:significance}, a maximum deviation at around 20 GeV is observed from all references with significance levels of 2 to 5$\sigma$, which is a characteristic feature of proposed signature of the QCD critical point predicted in Ref.~\cite{Stephanov:2011pb}.

At lower collision energy, on the other hand, changes are also see in proton factorial cumulant ratios, see panels (d), (e), and (f) of Fig.~\ref{fig:BESII_C4}. The values of both $\kappa_2/\kappa_1$ and $C_2/\langle p+\bar{p}\rangle$ are monotonically decreasing as collision energy decreases from 200 GeV to about 10 GeV primarily due to the baryon number conservation in the central Au+Au collisions. Below 10 GeV, however, the decreasing trend turns to increase in the lower energy collisions, $i.e.$, high baryon density region. The change in coverture in the energy dependence is also seen in peripheral collisions, open diamonds in Fig.~\ref{fig:BESII_C4}, although it appears not as strong as in the central collisions. While these decreasing energy dependence are reproduced by most of the model calculations without criticality, these calculations all failed to reproduce the increase in the ratios at lower energies. The change in the energy dependence may indicate the nature of interactions, namely, strongly repulsive interactions at the high energy end while attractive interactions become necessary at the high baryon density region~\cite{Friman:2025swg}. Note that in case of Van der Waals interaction, both attractive and repulsive interactions are required in order to form the critical point.

\begin{figure}[htb]
\includegraphics[width=0.99\textwidth]{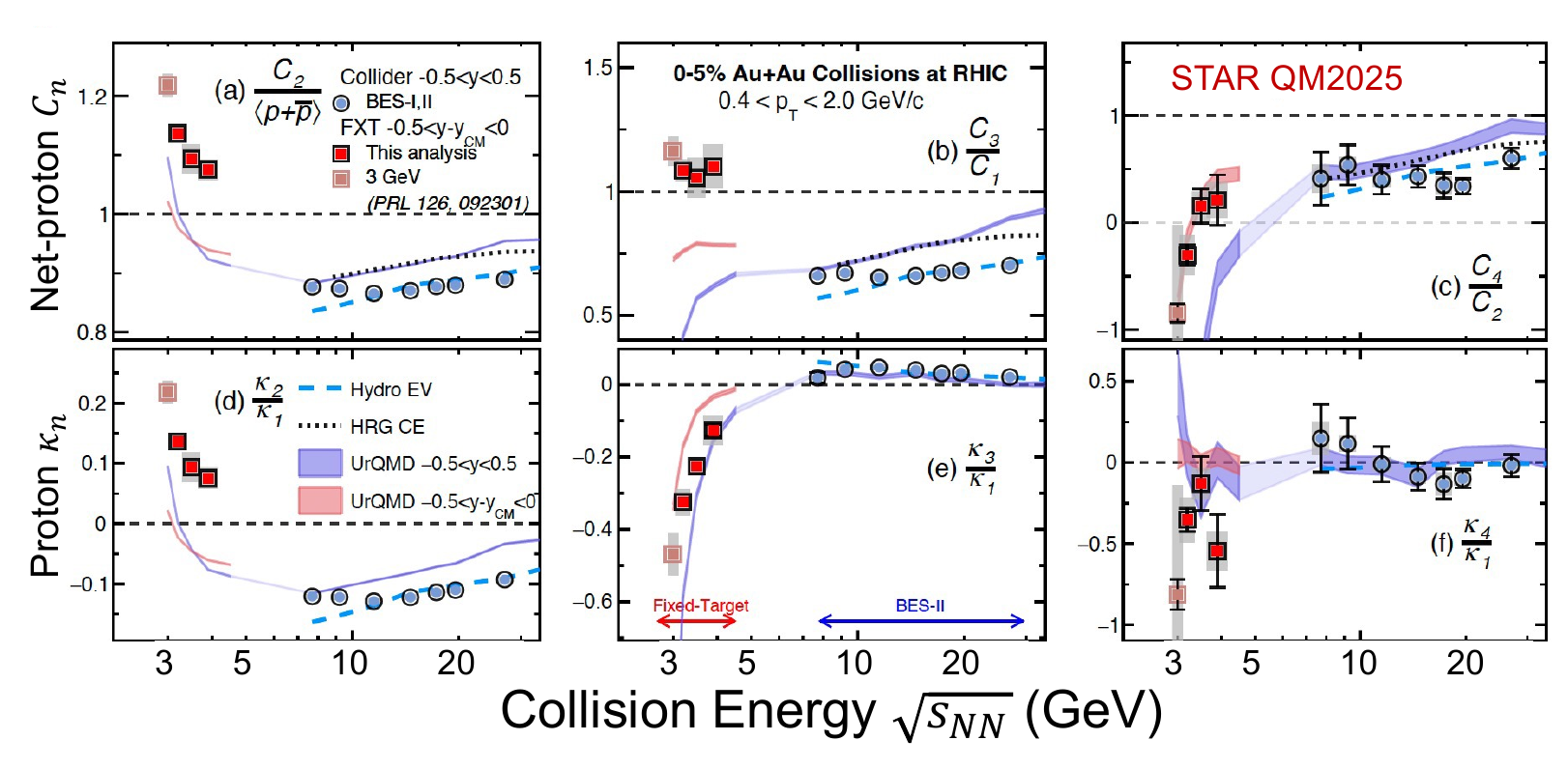}
\centering
\caption{Collision energy dependence of ratios for the cumulants of net-proton (top panels) and the factorial cumulant of protons (lower panels) in Au+Au collisions from STAR at RHIC. The red squares in the energy range $3<\sqrt{s_{\rm NN}} < 3.9$ GeV are the new results shown at Quark Matter 2025.}
\label{fig:FXT}
\end{figure}

Recent measurements on STAR fixed-target datasets in Fig.~\ref{fig:FXT} are shown at Quark Matter 2025 conference~\cite{Zach:QM2025}, which extend the energy dependence of proton cumulant and factorial cumulant ratios down to 3 GeV. In Fig.~\ref{fig:FXT} the new measurements in 0-5\% most central Au+Au collisions at $\sqrt{s_{\rm NN}}=3.0-3.9$ GeV are shown with red squares while results from collider data are shown with blue circles. Those measurements are performed with the rapidity window $-0.5<y<0$ in the center-of-mass frame and transverse momentum range $0.4<p_{\rm T}<2.0$ GeV/$c$. In panel (c) of Fig.~\ref{fig:FXT}, $C_4/C_2$ shows monotonically decreasing trend from collider energy to fixed-target energy due to baryon number conservation. While in panel (a) and (b), $C_2/\langle p+\bar{p}\rangle$ and $C_3/C_1$ shoot up at fixed-target energies which could indicate attractive interactions of particles~\cite{Friman:2025swg}. It is worth noting that at low collision energies, although the centrality bin width correction (CBWC) is used to suppress the volume fluctuations,  the limited centrality resolution, due to the smaller reference multiplicity, introduces significant volume fluctuation effects into the measurements. Therefore, caution should be exercised when interpreting the data. Recently, a centrality-independent method has been developed to address this issue~\cite{Wang:2025fve}, which will be discussed later in this review.  
The UrQMD calculations in Fig.~\ref{fig:FXT} are done within rapidity window $-0.5<y<0$ (blue bands) and $|y|<0.5$ (red bands), with centrality classes determined by a reference multiplicity distribution identical to that used in the data. In Fig.~\ref{fig:FXT}, the particle yields in UrQMD are not filtered by detector efficiency, resulting in a larger reference multiplicity, improved centrality resolution, and reduced volume fluctuation effects in the simulation. For a fair comparison with data, the UrQMD reference multiplicity should be scaled to match that of the experimental measurement. To suppress the initial volume fluctuation effect, a new method is proposed and discussed in Sec.~\ref{sec:ivf}. Currently, the measurements on STAR fixed-target datasets are in a preliminary stage. The recent result~\cite{Sorensen:2024mry} of the finite-size scaling analysis shows that the critical point is likely to exist in high baryon density region, which calls for final result at $\sqrt{s_{\rm NN}} = 4.5$ GeV. It should be emphasized that the subtraction of non-critical baselines assumes that critical contributions are separable from other dynamical effects. In a finite and rapidly evolving system, however, critical fluctuations are intertwined with finite-size effects, baryon transport, and acceptance dependencies. Therefore, observed deviations from baselines, while suggestive, cannot be uniquely attributed to criticality without further consistency checks.

\subsection{Suppression of initial volume fluctuation : A centrality independent method}\label{sec:ivf}
\begin{figure}[htb]
\centering 
\includegraphics[width=1.\textwidth] {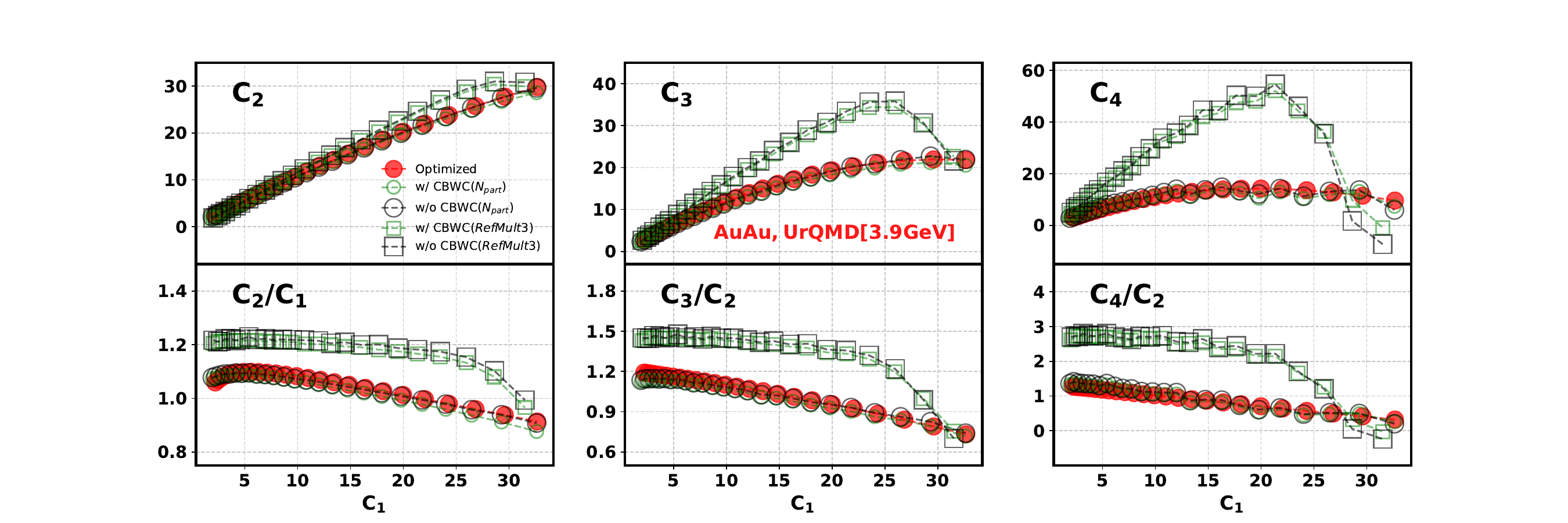}
\caption{Cumulants optimized from IVF correction method in UrQMD at $\sqrt{s_{\rm NN}} = 3.9$ GeV. Here black squares are from \emph{RefMult3}-based centrality without Centrality Bin Width Correction(CBWC), green squares are from \emph{RefMult3}-based centrality with CBWC, black circles are from $N_{\rm part}$-based centrality without CBWC, green circles are from \emph{$N_{\rm part}$}-based centrality with CBWC.}
\label{fig:Optimized_Results}
\end{figure}

Initial Volume fluctuations (IVF) and their impact on final-state particle multiplicity fluctuations in heavy-ion collisions have been studied over the years~\cite{Skokov:2012ds, Luo:2013bmi, Chatterjee:2019fey, Pawlowska:2021sea, Holzmann:2024wyd}. Experimentally defined centrality are based on uncorrected charged-particle multiplicity measurements. This definition does not strictly correspond to the geometric centrality, leading to the so-called volume fluctuation (VF) effect. These volume fluctuations have a strong impact on higher-order cumulants. Therefore, separating or suppressing the influence of volume fluctuations is crucial for accurate analysis of proton number fluctuations.
Figure \ref{fig:Optimized_Results} shows results at $\sqrt{s_{\rm NN}} = 3.9$ GeV from UrQMD which compares with that based on \emph{RefMult3}, here \emph{RefMult3} denotes the charged particle multiplicity by excluding protons - a design intended to eliminate auto-correlation effects in the analysis. It can be seen that IVF effect will affect the results of mid-central collisions in the case of \emph{RefMult3}-based centrality, and even for the most central collisions, the effect is non-negligible.
Recently, we have proposed a novel approach to suppress initial volume fluctuation effects~\cite{Wang:2025fve}. The methodology consists of three key components: (1) parameterizing the cumulants as polynomial functions of $C_1$, (2) ensuring the weights of proton distribution in each centrality bin are relatively close to each other, (3) reconstructing proton distributions for each centrality bin by substituting cumulant information into the Edgeworth expansion. To optimize this framework, we employ a differential evolution algorithm integrated with Bayesian optimization to systematically search for parameters that yield reconstructed total proton distributions in close agreement with data. This method has also been tested in UrQMD, and the results based on this approach are close to those obtained using \emph{$N_{\rm part}$}-based centrality.

Due to the issues with detector efficiency in the experiment, we conducted specific tests related to this problem. We tested the performance of the model regarding detector efficiency in UrQMD by simulating detector efficiency. We used the average efficiency $\epsilon$ dependent on centrality for correction, with the specific conversion formula as follows:

\begin{align}
\begin{split}
    C_{1,corr}  =& \lambda_1C_{1,uncorr}, \\
    C_{2,corr}  =& \lambda_2C_{2,uncorr}  + (\lambda_1  -\lambda_2)C_{1,uncorr}, \\
    C_{3,corr}  =& \lambda_3C_{3,uncorr}  + (3\lambda_2 -3\lambda_3)C_{2,uncorr}  + (\lambda_1  +2\lambda_3 -3\lambda_2)C_{1,uncorr}, \\
    C_{4,corr}  =& \lambda_4C_{4,uncorr}  + (6\lambda_3 -6\lambda_4)C_{3,uncorr}  + (7\lambda_2 - 18\lambda_3 + 11\lambda_4)C_{2,uncorr}  + \\
                &(\lambda_1  - 7\lambda_2 + 12\lambda_3  - 6\lambda_4 )C_{1,uncorr}.
\end{split}
\end{align}
Here $\lambda_1=1/\epsilon, \lambda_2=1/\epsilon^2,   \lambda_3=1/\epsilon^3, \lambda_4=1/\epsilon^4$. 

\begin{figure}[htb]
\centering
\includegraphics[width=0.85\textwidth]{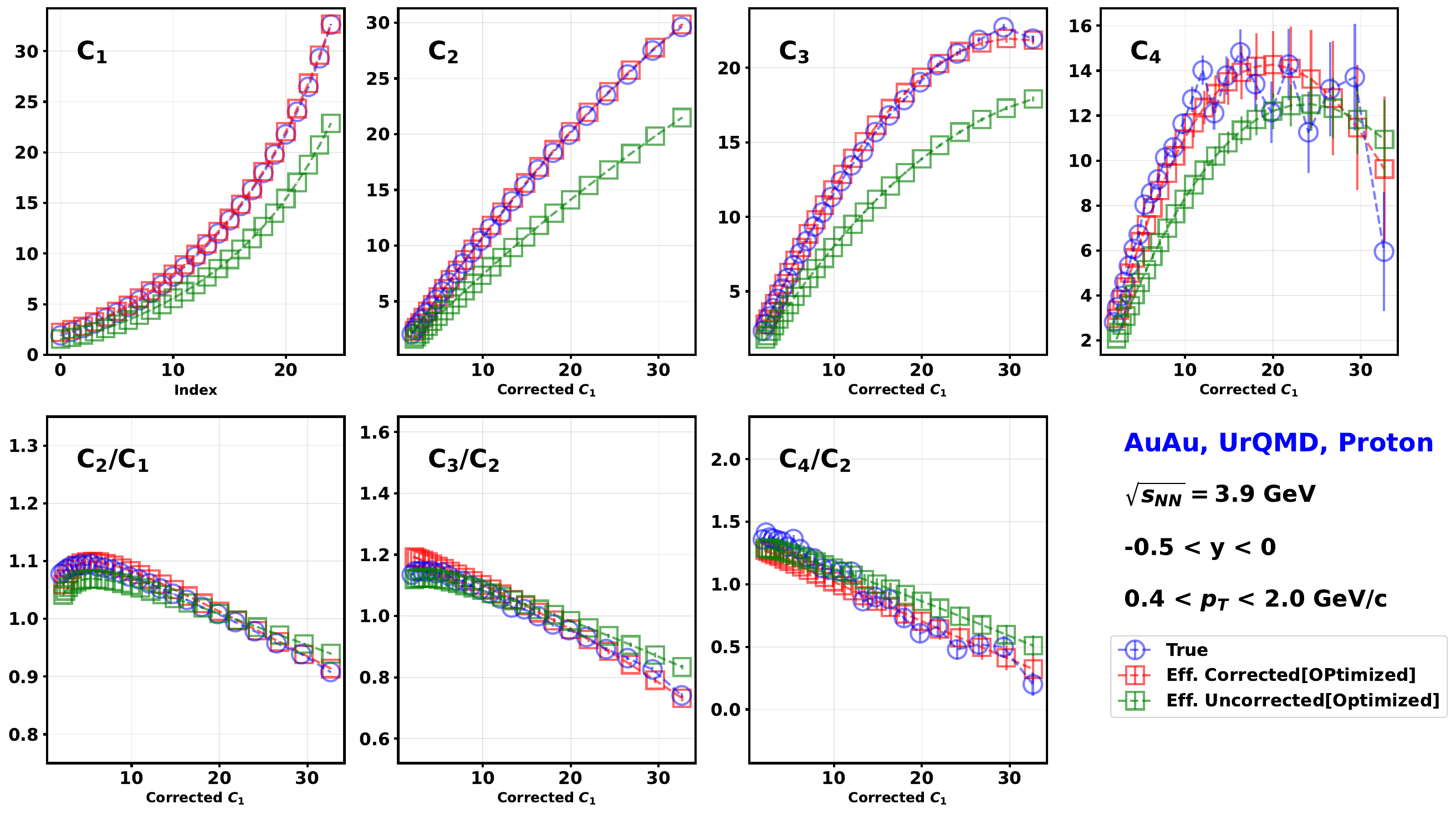}
\caption{Detector efficiency test in Au+Au collisions at $\sqrt{s_{\rm NN}} = 3.9$ GeV in UrQMD. Index here means centrality bin index, and the blue circles represent the results based on \emph{$N_{\rm part}$}-based centrality, while the green and red squares are the results before and after efficiency correction, respectively.}
\label{fig:Efficiency_Test}
\end{figure}

Figure \ref{fig:Efficiency_Test} shows detector efficiency test in Au+Au collisions at $\sqrt{s_{\rm NN}} = 3.9$ GeV in UrQMD. We applied an average efficiency of 0.7 to the proton tracks and evaluated the applicability of our method. Here in this figure the blue circles represent the cumulants based on $N_{\rm part}$-based centrality, while the green and red squares are the results obtained using the afore-mentioned method before and after efficiency correction, respectively. Comparing the red squares with the blue circles, we can see they are close, demonstrating the validity of our efficiency correction method.

\section{Summary and outlook}\label{sec:summary}
In summary, we review recent precision measurements on (net-)proton cumulant and factorial cumulant ratios in Au+Au collisions at $\sqrt{s_{\rm NN}} = 3-200$ GeV by the STAR detector at RHIC. A maximum deviation of net-proton cumulant ratio $C_4/C_2$ from baseline calculations without implementing a QCD critical point is observed in 0-5\% central Au+Au collisions at $\sqrt{s_{\rm NN}}=19.6$ GeV. On the other hand, at low energies especially $\sqrt{s_{\rm NN}} < 11$ GeV, lower order ratios, net-proton $C_2/\langle p+\bar{p}\rangle$, $C_3/C_1$ and proton $\kappa_2/\kappa_1$, $\kappa_3/\kappa_1$, show clear deviations from baselines indicating the needs of attractive interactions at the high baryon density region, as discussed in Ref.~\cite{Friman:2025swg}. Measurements of proton cumulant ratio $C_{4}/C_{2}$ in 0-5\% central Au+Au collisions at $\sqrt{s_{\rm NN}}=3-3.9$ GeV seem to follow the decreasing trend from high energies. While these observations are consistent with theoretical expectations for proximity to a QCD critical point, alternative explanations related to dynamical baryon transport, finite-size effects, and acceptance dependencies cannot be ruled out. The interpretation of data is complicated by the finite and non-equilibrium nature of heavy-ion collisions. To address these challenges, we have discussed a new centrality-independent method to suppress initial volume fluctuation effects, which is particularly important for low-energy fixed-target data. Furthermore, the community is increasingly exploring finite-size scaling approach and other dynamical frameworks to more robustly search for QCD critical point.

Looking ahead, the final results from the STAR fixed-target program ($\sqrt{s_{\mathrm{NN}}} = 3 - 4.5$ GeV) will provide essential data in the high baryon density region. Beyond RHIC, there are several accelerator complexes which could also carry out the physics analysis for the QCD critical point. On papers, both Compressed Baryonic Matter experiment (CBM) at FAIR~\cite{CBM:2016kpk} and Multi-Purpose Detector experiment (MPD) at NICA~\cite{Kekelidze:2018fvh} will be ready to take data around the end of 2028. The combined energy coverage from these two facilities is 2.0 $\le \sqrt{s_{\rm NN}} \le$ 11.2 GeV, overlaps well with the RHIC BES program, especially at the high baryon density region. These new generation of experiments with high intensity capability will improve the quality of data both statistically and systematically. In addition, the new accelerator High Intensity Heavy-ion Facility (HIAF)~\cite{Zhou:2022pxl} under construction in Huizhou, China, is in its early commissioning period. This machine will offer $p=9.3$ GeV/$c$ proton beams and 2.5 GeV/$c$ U beam with the intensity of $10^{11}$ and of $10^{9}$, respectively.  
Scientists are considering a new spectrometer that will focus on the measurements of baryon polarizations, multiplicity distributions and correlations for the physics of hadron spin, critical fluctuations, and hyperon-nucleon interactions, respectively.

\section{Acknowledgments}\label{sec:ack}
The authors would like to thank Drs. P. Braun-Munzinger, H.T. Ding, X. Dong, W.J. Fu, F. Gao, F. Karsch, V. Koch, B. Monhaty, J. Pawlowski, K. Redlich, A. Rustamov, J. Stachel, M. Stephanov for insightful discussions. This work was supported by National Natural Science Foundation of China (No. 12525509 and 12447102), National Key Research and Development Program of China (No.2022YFA1604900). Yu Zhang was supported by the Guangxi Natural Science Foundation (No. 2025GXNSFBA069067).

\section*{Data Availability Statement}
This paper is a review and does not report new primary data. All data discussed are available from the published sources cited in the reference list. The original data from the STAR experiment are subject to the collaboration's data policy and are typically made publicly available through the RHIC/STAR Data Portal (https://www.star.bnl.gov). Processed data points for specific figures are often available within the corresponding published articles or via contact with the authors of those primary publications.

\bibliography{reference}

\end{document}